\documentclass[11pt,aps,prd,superscriptaddress,amsmath,amssymb,nofootinbib]{revtex4-2}
\usepackage[T1]{fontenc}
\usepackage[utf8]{inputenc}
\usepackage[colorlinks=true, allcolors=blue]{hyperref}

\begin{document}

\title{Fully conservative $f(R,T)$ gravity and Solar System constraints}
\author{Nicolas R. Bertini}\email{nicolas.bertini@ufop.edu.br}
\affiliation{Departamento de Ciências Exatas e Aplicadas, Universidade 
Federal de Ouro Preto, Campus João Monlevade, João Monlevade, MG, Brazil}
\author{Hermano Velten}\email{hermano.velten@ufop.edu.br}
\affiliation{Departamento de Física, Universidade 
Federal de Ouro Preto, Campus Morro do Cruzeiro, Ouro Preto, MG, Brazil}

\begin{abstract}
The $f(R,T)$ gravity is a model whose action contains an arbitrary function of the Ricci scalar $R$ and the trace of the energy-momentum tensor $T$. We consider the minimally coupled model $f (R, T ) = \chi(R) + \varphi(T )$ and shown that, for perfect fluids, the analysis of dynamical equations are sufficient to determine how $\varphi$ depends on $T$, independently of the matter fields equation of state and the geometry of space-time. Imposing the energy-momentum tensor conservation we obtain that the trace dependent part $\varphi$ must be linear in $T$, apart from the trivial case of a constant. However, the linear dependence on $T$ is severely constrained using the full Will-Nordtvedt version of the parameterized post-Newtonian (PPN) formalism. The result of the PPN analysis is discussed and in addition it is shown that the diffeomorphism invariance of the matter action imposes strong constraints on conservative versions of $f(R,T)$ gravity.
\end{abstract}

\maketitle

\section{Introduction}

General relativity is considered the correct description for the gravitational interaction due to its huge success in explaining available astronomical data at the solar system level and also for first predicting new gravitational phenomena posteriorly observed. On the other hand, the existence of the dark sector of the universe, consisting of dark matter and dark energy, is one of the most intriguing scientific challenges of the last decades and remains unsolved. By trusting that general relativity is the correct theory for describing the gravitational interaction, specially above the galactic scales, the standard cosmological model demands that about $25\%$ of current cosmic energy budget should be composed by dark matter while other $70\%$ sums up the dark energy sector. So far, though there exists a large number of theoretically well motivated particles/fields candidates to compose the dark sector, there is no direct detection of them. Such lack of detection has motivated the search for a new theory in replacement of GR. Currently, this is a vast research field \cite{Clifton:2011jh, CANTATA:2021ktz}.

Either by modifying GR structures or by just adding new elements that yield to extended GR versions, all existing proposals should be subjected to the fact that there are no reasons to doubt GR validity at local scales i.e., the solar system one. Apart from analytical and numerical results for the behavior of a gravitational theory at the solar systems level, this analysis can also be technically performed via the so called post-Newtonian formalism \cite{Will:1993ns, will2014confrontation,blanchet2014gravitational}.

In this work we shall be concentrated on a widely known modified theory of gravity in which the gravitational action of the total theory depends on the trace of the energy momentum $T\equiv T^{\mu}_{\mu}$ in addition to the Ricci scalar $R$ \cite{Harko:2011kv}. In short notation, starting with $f(R,T)$ gravity, GR is fully recovered with $f(R,T)\equiv R$.
In the literature there are many distinct formulations for the functional dependence of $f(R,T)$. For example, the $T$ dependence has been written either in the polynomial, rational or the exponential one \cite{Yousaf:2016lls, Shabani:2014xvi, Shabani:2013djy, Moraes:2015uxq, Alvarenga:2012bt, Azizi:2012yv, Roshan:2016mbt, Noureen:2015nja, Moraes:2019hgx}. 

When seeking to determine the functional form of $f(R,T)$, it is not strictly required but highly desirable for the ensuing dynamics to exhibit conservative behavior, i.e., preserve the vanishing of the four-divergence of the energy momentum tensor, $\nabla^{\mu}T_{\mu\nu}=0$. Maintaining this property ensures that the underlying physical system is conservative which is a fundamental requirement for a wide range of physical phenomena but, specially for applying the post-Newtonian formalism. Therefore, incorporating this criterion into the search for the function $f(R,T)$ is a key factor. For example, cosmological solutions have shown that energy-density conservation can be reinstated in a manner similar to general relativity (GR) if $f(R,T) = f_1(R)+f_2(T)$, where $f_2$ is defined as $A+BT^{1/2}$ and $A$ and $B$ are constants. \cite{Alvarenga:2013syu}. The issue of conservative $f(R,T)$ models have also been explored in the context of spherically symmetric stellar configurations. In such analysis, it is demanded that the effective energy momentum tensor should be conserved, establishing a criteria for the allowed polytropic equation of state parameters \cite{dosSantos:2018nmu, Pretel:2021kgl}. Though the possibility that conservation can be violated, in the sense that $\nabla^{\mu}T_{\mu\nu} \neq 0$, the entire mechanism behind the post-Newtonian analysis demands conservation and has, as we will show below, a strong constraining power on the functional form of $f(R,T)$. This is the main result of this work.

At the level of cosmological solutions, the viability of $f(R,T)$ theories on cosmological scales has been addressed in the literature. Though it has been argued that the high redshift evolution of $f(R,T)$ cosmologies are incompatible with standard cosmological model \cite{Velten:2017hhf}, a more complete recent analysis using the age of the universe as a constraining cosmological observable has proved that there is however a small free  parameter space $f(R,T)$ model that can lead to viable cosmological scenarios at the background level \cite{Jeakel:2023xlp}.

Our goal in this work is to investigate the general feature of conservation in $f(R,T)$ theories at the level of the field equations, i.e., before setting a specific geometry. We show that the case $f_2 \propto T^{1/2}$ represents a particular ``cosmological'' case of the full conservative $f(R,T)$ theory. In addition, by applying the parameterized post-Newtonian (PPN) formalism we demonstrate that the existence of conservative $f(R,T)$ theories is challenged. The only exception is the trivial cosmological constant like case.

This work is organized as follows. In the next section we review the field equations for $f(R,T)$ gravity. In the third section we focus on discussing the role played by a minimally coupled $f(R,T)= f_1(R)+f_2(T)$ choice. We then work out the full conservative model at the field equations level and present one of the main results of this work. In Sec. IV we briefly review some essential PPN features and apply them to the case of minimally coupled $f(R,T)$ gravity. In Sec. V the results from the previous section are discussed and the observational bounds for the case developed in Sec. III are determined. In Sec. VI we consider the action invariance under diffeomorphisms and find the condition that must be satisfied for a general $f(R,T)$ theory to be conservative. Our conclusions are in Sec. VII.

\section{The action and field equations}
 
The $f(R,T)$ gravity was first proposed in \cite{Harko:2011kv}, and characterized  by an arbitrary dependence of the gravitational action on the Ricci scalar and the energy-momentum tensor trace $T = T^{\mu}_{\;\;\mu}$. The full action is 
\begin{align}
 S[g,\Psi] =  \frac{1}{2\kappa^{2}}\int f(R,T) \sqrt{-g}\,d^{4}x + S_\mathrm{m}[g,\Psi],\label{eq:action}
\end{align}
where $\kappa^{2} = 8\pi G$ being $G$ the gravitational coupling constant ($c=1$), $f(R,T)$ is an arbitrary function of the $R$ and $T$,  $S_\mathrm{m}$ is the action for the matter fields denoted collectively by $\Psi$. The energy-momentum tensor is obtained from $S_\mathrm{m}$ by the usual definition
\begin{align}
 T_{\mu\nu}\equiv -\frac{2}{\sqrt{-g}}\frac{\delta S_\mathrm{m}}{\delta g^{\mu\nu}}.\label{eq:TEM}
\end{align}

The variation of action \eqref{eq:action} with respect to the metric results in 
\begin{align}
 f_{_R}R_{\mu\nu} - \frac{1}{2}g_{\mu\nu}f + (g_{\mu\nu}\Box - \nabla_{\mu}\nabla_{\nu})f_{_R} = \kappa^{2} T_{\mu\nu} - f_{_T}(T_{\mu\nu} + \Theta_{\mu\nu}),\label{eq:FE1}
\end{align}
where $f\equiv f(R,T)$ and, for sake of simplicity, we defined
\begin{align}
f_{_R} \equiv \frac{\partial f(R,T)}{\partial R}, \quad  f_{_T} \equiv \frac{\partial f(R,T)}{\partial T} \quad\mathrm{and}\quad  \Theta_{\mu\nu} \equiv g^{\alpha\beta}\frac{\delta T_{\alpha\beta}}{\delta g^{\mu\nu}}.
\end{align}
 The equation for the energy-momentum tensor divergence, obtained directly from \eqref{eq:FE1}, is
\begin{align}
\nabla^{\mu}T_{\mu\nu} = \frac{f_{_T}}{\kappa^{2} - f_{_T}}\left[ -\frac{1}{2}\nabla_{\nu}T + (T_{\mu\nu}+\Theta_{\mu\nu})\nabla^{\mu}\ln f_{_T} + \nabla^{\mu}\Theta_{\mu\nu} \right].\label{eq:09}
\end{align}
Of course, $f_T=0$ restores the conservation. 
The energy momentum tensor for perfect fluids reads
\begin{eqnarray}
T_{\mu\nu} = p (u_{\mu}u_{\nu}+g_{\mu\nu})+  \varepsilon u_{\mu}u_{\nu},\label{eq:TEM1}
\end{eqnarray}
where $p$ is the pressure, $\varepsilon$ is the energy density and $u^{\alpha}$ is the four velocity of the fluid element. In this case
\begin{align}
\Theta_{\mu\nu} = -2T_{\mu\nu} +g_{\mu\nu}p,
\end{align}
and $T$ depends on the energy density and pressure through
\begin{align}
T = -\varepsilon+3p.\label{eq:T}
\end{align}
Furthermore, the matter field dynamics is related to $f$ by the equation
\begin{align}
 \frac{1}{2\kappa^{2}}\int d^{4}x'\sqrt{-g'}\frac{\delta f'}{\delta \Psi} = -\frac{\delta S_\mathrm{m}}{\delta\Psi},\label{eq:matt}
\end{align}
where $x'$ denotes the integration variable. Therefore, it should be clear that $p$ and $\varepsilon$ in above equations depend on $f$, by construction.

\section{The case of a minimal coupling}

This section outlines the fundamental characteristics of the gravitational field equations when expressing the function $f(R,T)$ in the separable form of $f_1(R)+f_2(T)$. Here, the matter sector, as given by the trace $T$, is minimally coupled with the geometric quantities found in the gravitational Lagrangian. It is noteworthy that in this case, the conservation of the energy momentum tensor is essential for successful application of the PPN formalism, and such conservation is indeed observed. However, it is worth mentioning that nonminimally coupled scenarios are intrinsically nonconservative \cite{Jeakel:2023xlp}.

Imposing both the vanishing of the four-divergence of $T_{\mu\nu}$ in Eq.~\eqref{eq:09} i.e., $\nabla^{\mu}T_{\mu\nu}=0$ and $f_{_{T}}-\kappa^{2}\neq 0$, and also taking into account that $\nabla^{\mu}\Theta_{\mu\nu} = \nabla^{\mu}(-2T_{\mu\nu} + g_{\mu\nu}p)=\nabla_{\nu}p$,  this yields to the constraint
\begin{eqnarray}
f_{_T}\left[ -\frac{1}{2}\nabla_{\nu}T + (g_{\mu\nu}p - T_{\mu\nu})\nabla^{\mu}\ln f_{_T} + \nabla_{\nu}p \right]=0.
\end{eqnarray}
Since $f_{_{T}}\nabla^{\mu}\ln f_{_{T}} = \nabla^{\mu}f_{_{T}}$ one has
\begin{eqnarray}
 -\frac{1}{2}(\nabla_{\nu}T) f_{_T} + (g_{\mu\nu}p - T_{\mu\nu})\nabla^{\mu} f_{_T} + (\nabla_{\nu}p) f_{_T}=0. \label{eq:4}
\end{eqnarray}
From this point one has to make a choice on the $f(R,T)$ function. Let us consider the minimal coupling case
\begin{eqnarray}\label{eqfRT}
f(R,T) = \chi (R)+\varphi (T).
\end{eqnarray}

According to Eq.~\eqref{eq:matt} it is technically possible to assume the dependence of $p$ on $\varepsilon$ and $\varphi(T)$. For any equation of state of the type $p=p(\varepsilon)$ admitting an inverse $\varepsilon=\varepsilon(p)$ one has
$T =  -\varepsilon(p) + 3p$, and therefore it is possible to write $p = p(T)$. This is an important assumption made in our work.

Now, let $\zeta \equiv\zeta(R,T)$ be any function of $R$ and $T$. Then, without loss of generality, the four-divergence of this quantity is directly computed as $\nabla_{\mu}\zeta = \nabla_{\mu}T \zeta_{_T} + \nabla_{\mu}R \zeta_{_R}$. Therefore, it is trivial to write down the quantities $\nabla_{\mu}f_{_T}=(\nabla_{\mu}T)f_{_{TT}}+(\nabla_{\mu}R)f_{_{RT}} = (\nabla_{\mu}T)\varphi_{_{TT}}$ and $\nabla_{\mu}p = (\nabla_{\mu}T)p_{_T}$. This allows one to rewrite \eqref{eq:4} into the form
\begin{eqnarray}
\left[ (g_{\mu\nu}p-T_{\mu\nu})\varphi_{_{TT}} + \frac{1}{2}g_{\mu\nu}(2p_{_T} -1)\varphi_{_T} \right]\nabla^{\mu}T=0.
\end{eqnarray}
Therefore, in general we have 
 \begin{align}
 (p\delta^{\mu}_{\nu}-T^{\mu}_{\nu})\varphi_{_{TT}} + \frac{1}{2}(2 p_{_T} -1)\varphi_{_T}\delta^{\mu}_{\nu}=0.\label{eq:phi}
 \end{align}

The $\mu=\nu=0$ and $\mu=i,\;\;\nu=j$ components of above equation are (in the comoving frame, $u^{i}=0$), respectively
\begin{align}
(p+\varepsilon)\varphi_{_{TT}}+ \left(p_{_T} - \frac{1}{2} \right)\varphi_{_T}=0,\label{eq:sol100}
\end{align}
\begin{align}
\left(p_{_T} - \frac{1}{2} \right)\varphi_{_T}\delta^{i}_{j}=0.\label{eq:sol1ij}
\end{align}

In the cosmological context, considering the state equation $p = w\varepsilon$, the solution obtained in the Refs.~\cite{Alvarenga:2013syu, Baffou:2013dpa} is consistent with that found considering only the Eq.~\eqref{eq:sol100}: $\varphi(T) = A T^{\frac{1+3w}{2(1+w)}} + B$, for $w\neq 1/3$ and $A$ and $B$ being integration constants. As argued in \cite{Alvarenga:2013syu} this model should represent the only viable $f (R,T)$ theory, since it constitutes the only case in which the standard conservation law is preserved. However, the viability of this model in explaining cosmology background observables is discussed in \cite{Velten:2017hhf} and the findings of such reference represent a challenge to this model as a viable modification of gravity. However the approach in \cite{Alvarenga:2013syu, Baffou:2013dpa} is such that Eq.~\eqref{eq:sol1ij} is not considered.

The solution of Eq.~\eqref{eq:sol100} using the condition \eqref{eq:sol1ij}, even if one does not know the equation of state, is
\begin{align}
\varphi(T) = \sigma_0 + \sigma_1 T ,\label{eq:h(T)}\quad (\,p\neq - \varepsilon\,)
\end{align}
being $\sigma_0$ end $\sigma_1$ constants. The only condition behind the above result is $p\neq - \varepsilon$.

Concerning Eqs.~\eqref{eq:sol100} and \eqref{eq:sol1ij} it is worth noting that, excluding the case where $\varphi_{_T}=0$, the choice of the comoving frame implies a relationship between $p$ and $T$ so that, given an equation of state $p=p(\varepsilon)$, the fluid equation of state parameter can be determined. For example, assuming $p=\omega\varepsilon$, Eq.~\eqref{eq:sol1ij} implies $\omega = 1$, the stiff matter model.

On the other hand, the solution \eqref{eq:h(T)} is found for any arbitrary reference frame. To verify this one simply takes the trace of Eq.~\eqref{eq:phi} and uses the result to eliminate $(2p_{_T}-1)\varphi_{_T}$. This procedure results in
\begin{equation}
    \left(T_{\mu\nu}-\frac{1}{4}g_{\mu\nu}T\right)\varphi_{_{TT}}=0\,.
\end{equation}
The above equation can be satisfied in two cases: (i) if $\varphi (T)$ is given by \eqref{eq:h(T)}; or (ii) if the term in parentheses is zero. The second case occurs, in general, if $p=-\varepsilon$. This relation corresponds to the general relativistic solution for a cosmological constant which is equivalent to the case $\varphi = const$. However, the relation \eqref{eq:h(T)} is even more general by including this case.

Since we have determined a functional form for $f(R,T)$ that leads to a conservative energy-momentum tensor, in the next section we apply the full Will-Nordevedt PPN formalism to this model and restrict the $\sigma$'s parameters of the Eq.~\eqref{eq:h(T)} in section \ref{sec:V}. For this we specify the functional form
\begin{eqnarray}
    f(R,T) = R + \varphi(T),\label{eq:min}
\end{eqnarray}
which corresponds to $\chi(R) = R$ in Eq.\eqref{eqfRT}, and 
\begin{align}
    \varphi(T) \equiv \varphi_n(T) = \sum^{n}_{i=0}\sigma_{i}T^{i}.\label{eq:phin}
\end{align}
The solution \eqref{eq:h(T)} is obtained for n=1. Substituting into the field equation \eqref{eq:FE1} we have
\begin{align}
 R_{\mu\nu} - \frac{1}{2}g_{\mu\nu}R =  \frac{1}{2}g_{\mu\nu}\varphi  + 8\pi G T_{\mu\nu} - \varphi_{_T}(T_{\mu\nu} + \Theta_{\mu\nu}).\label{eq:16}
\end{align}
It is useful to write this expression in the form
\begin{align}
 R_{\mu\nu} = -\frac{1}{2}g_{\mu\nu} \varphi + 8\pi G \left( T_{\mu\nu} - \frac{1}{2}g_{\mu\nu}T \right) - \varphi_{_T} \left( T_{\mu\nu} + \Theta_{\mu\nu} - \frac{1}{2}g_{\mu\nu}(T+\Theta) \right),\label{eq:FE2}
\end{align}
where the trace was used to eliminate $R$.

\section{Post-Newtonian expansion}

In this section we apply the Will-Nordtvedt PPN formalism  \cite{Will:1993ns, Will:2014kxa} to $f(R,T)$ gravity in its minimally coupled form 
\eqref{eq:min}. In this formalism, a perfect fluid is the source of the gravitational field. It describes the metric of a gravitational theory in terms of ten observable PPN parameters in a theory-independent way. The main small parameter of this formalism is the matter velocity field $|\vec v| = v < 1$. The metric is expanded about Minkowski spacetime,
\begin{equation} \label{eq:getah}
g_{\alpha\beta}=\eta_{\alpha\beta}+h_{\alpha\beta}\,,
\end{equation}
where  $\eta_{\alpha\beta}$ is the Minkowski metric, which is of zeroth-order on $v$, and $h_{\alpha \beta} \sim O(v^2)$, at least. We use the signature $(-,+,+,+)$.

Up to the first post-Newtonian order, the metric must be known as follows: $g_{00}$ to order $v^4$, $g_{0i}$ to order $v^3$ and $g_{ij}$ to order $v^2$ (Latin indices run from $1$ to $3$). Thus, up to the required order, the Ricci tensor components can be expressed as 
\begin{align}
R_{00} =& -\frac{1}{2}\nabla^2h_{00} - \frac{1}{2}\left(h^k_{~k,00}- 2\,h^k_{~0,k0} \right) - \frac{1}{4}\,|\vec{\nabla}h_{00}|^2 \ +\nonumber\\[1ex]
&+\frac{1}{2}\,h_{00,l}\left(h^{lk}_{~~,k}- \frac{1}{2}\,h^k_{~k,j}\delta^j_l\right)+ \frac{1}{2}\,h^{kl}h_{00,lk}\,,\label{1}\\[2ex]
R_{0i}=& -\frac{1}{2}\left(\nabla^2h_{0i} - h^k_{~0,ik} + h^k_{~k,0i} - h^k_{~i,k0} \right)\,,\label{2}\\[2ex]
R_{ij}=& -\frac{1}{2}\!\left(\nabla^2h_{ij} -\! h_{00,ij} +\! h^k_{~k,ij}-\! h^k_{~i,kj}- \! h^k_{~j,ki} \right).\label{3}
\end{align}
The comas refer to partial derivatives, $\nabla^2 \equiv \eta^{ij}\partial_i\partial_j$, and it is used that time derivatives effectively yield to a higher order in the expansion. Thus, if a quantity $X$ is of order $v^n$ then $X,_k\sim O(v^n)$ and $X,_0\sim O(v^{n+1})$.

Since the gravitational source is  a perfect fluid, the energy-momentum tensor is the one given in \eqref{eq:TEM1} but the energy density is decomposed into the mass density $\rho$ and the specific energy density $\Pi$ in the form
\begin{equation}\label{emt}
\varepsilon = \rho+\rho\Pi.
\end{equation}
 The four velocity of the fluid element $u^\mu=(u^0,v^i)$, with
\begin{equation}\label{u0}
u^0= \sqrt{\frac{1+v^2}{1-h_{00}}}\,,
\end{equation}
such that $u^\mu u_\mu=-1$. The density $\rho$, $\Pi$ and $p/\rho$ are of order $v^2$ \cite{Will:1993ns}.

With the expressions above, we expand Eq.~\eqref{eq:FE2} to calculate the metric components order by order on powers of $v$.  As a first step, the zeroth order equation in $v$ trivially leads to 
\begin{equation}
    \sigma_{0}=0.
\end{equation}
This is expected since the constant $\sigma_{0}$ in Eq.~\eqref{eq:h(T)} necessarily leads to nonasymptotically flat spacetimes, like the cosmological constant in general relativity, which is not considered in the standard PPN Solar System analysis. This is also physically reasonable since, up to first post-Newtonian order (1PN) and considering its value as inferred from the cosmological observations, it has negligible impact on the Solar System dynamics \cite{Sereno:2006re}. The next steps are described below.
\begin{itemize}
\item{$h_{00}$ up to order $v^2$ (Newtonian limit):}
Up to the required order one finds
\begin{equation}
R_{00}=-\frac{1}{2}\,\nabla^2h_{00},\;\; \varphi_{1} = \sigma_{1}T, \;\; T_{00}=-T=\rho\, \quad \mbox{and} \quad \Theta_{00} = -\Theta = 2\rho.
\end{equation}
Inserting this into Eq.~\eqref{eq:FE2} we find that the terms containing $\sigma_{1}$ cancel out, therefore
\begin{equation}
    \nabla^{2}h_{00}= - 8\pi G\rho\,.
\end{equation}
In order to be in agreement with the local Newtonian gravity it is demanded then,
\begin{equation}
    h_{00}=2U,
\end{equation}
where $U$, the negative of the Newtonian potential\footnote{For conciseness, commonly we will call $U$ the Newtonian potential, without writing ``negative'' in front of it. We use $U$ since we are following the notation of Ref.~\cite{Will:1993ns} on the PPN parameters and the potentials.} \cite{Will:1993ns}, given by
\begin{equation}
    U(t, \vec{x}) = G_\mathrm{N} \int \frac{\rho(t, \vec{x} \, ')}{| \vec{x} - \vec{x} \, '| } d^3x' \, ,
\end{equation}
being $G_\mathrm{N}$ the Newtonian gravitational constant. Thus we must set
\begin{equation}
    G=G_\mathrm{N}=1.
\end{equation}

\item{$h_{ij}$ up to order $v^2$:}
Imposing the three gauge conditions, 
\begin{equation}
    h^{\mu}_{\;i,\mu} - \frac{1}{2} h^{\mu}_{\;\mu,i} = 0,\label{eq:gauge1}
\end{equation}
the spatial part of Eq.~\eqref{eq:FE2} reduces to
\begin{equation}
    -\frac{1}{2}\nabla^{2}h_{ij} = 4\pi\rho\delta_{ij} + \sigma_{1}\rho\delta_{ij}.
\end{equation}
This equation is easily integrated, providing
\begin{equation}
    h_{ij} = 2 \left( 1+ \frac{\sigma_1}{4\pi} \right)U\delta_{ij}.
\end{equation}

\item{$h_{0i}$ up to order $v^3$:}
With the fourth gauge condition
\begin{equation}
    h^{\mu}_{\;0,\mu} - \frac{1}{2}h^{\mu}_{\;\mu,0}=\frac{1}{2}h_{00,0}\,,\label{eq:gauge2}
\end{equation}
the Eq.~\eqref{eq:FE2} becomes
\begin{eqnarray}
\nabla^{2}h_{0j} + U_{,0j} = 16\pi\left(1+\frac{h_1}{8\pi} \right)\rho v_{j}.
\label{eq:Dh0j}
\end{eqnarray}
The above equation can be integrated using the auxiliary potential $\chi(t,\vec{x})$ \cite{Will:1993ns}, given by
\begin{equation}
    \chi (t,\vec{x})\equiv\int \rho(t,\vec{x}')|x-\vec{x}'|d^{3}x'.
\end{equation}
From this definition one can write
\begin{equation}
    \nabla^{2}\chi = -2U \quad \mathrm{and} \quad \chi_{,0j} = V_j - W_j,
\end{equation}
where
\begin{equation}
    V_j = \int \frac{\rho(x-\Vec{x}')v'_{i}}{|x-\vec{x}'|}d^{3}x'\,, \quad \nabla^{2}V_j = -4\pi\rho v_j\,.
\end{equation}
Therefore, Eq.~\eqref{eq:Dh0j} gives 
\begin{eqnarray}
h_{0j} = -\left( \frac{7}{2} +\frac{h_1}{2\pi} \right)V_j - \frac{1}{2}W_j\,.
\end{eqnarray}

\item{$h_{00}$ up to order $v^4$:}
To the required order
\begin{equation}
    \varphi_{2} = \sigma_{1}(-\rho +3p)+\sigma_{2}\rho^{2},
\end{equation}
and for the energy-momentum tensor, one finds
\begin{equation}
    T_{00} - \frac{1}{2}g_{00}T = \rho \left( v^{2} - U +\frac{\Pi}{2} + \frac{3p}{2\rho} \right).
\end{equation}
\end{itemize}
With the above relations the $0-0$ component of Eq.~\eqref{eq:FE2} gives
\begin{align}
R_{00} =  - 4\pi\rho v^{2}\left( 2+\frac{\sigma_1}{4\pi} \right)-4\pi\rho U\left( -2+\frac{\sigma_{1}}{\pi}\right) -4\pi\rho\Pi
 - 4\pi p\left(3 + \frac{\sigma_{1}}{2\pi} \right) + 4\pi\rho^{2}\frac{\sigma_{2}}{8\pi}.\label{eq:R00}
\end{align}
Let us now find the $0-0$ component of the Ricci tensor. Considering the post-Newtonian potentials introduced in \cite{Will:1993ns}
\begin{align}
\nabla^{2}\phi_{1}  = -4\pi\rho v^{2}, \quad \nabla^{2}\phi_{2}  = -4\pi \rho U,
\\
\nabla^{2}\phi_{3}  = -4\pi \rho\Pi, \quad \nabla^{2}\phi_{4}  = -4\pi p,
\end{align}
we take into account the gauge conditions \eqref{eq:gauge1} and \eqref{eq:gauge2}, and using the relation $|\vec{\nabla}U| = \nabla^{2}(U^{2}/2 - \phi_2)$, the Ricci tensor component reads
\begin{equation}
    R_{00} = -\frac{1}{2}\nabla^{2}(h_{00}+2U^{2}-8\phi_2).
\end{equation} 
With the above results, eq.~\eqref{eq:R00} becomes
\begin{eqnarray}
h_{00} = -2U^{2} + \left( 4+\frac{\sigma_{1}}{2\pi} \right)\phi_{1} + \left(4+\frac{2\sigma_{1}}{\pi} \right)\phi_{2} + 2\phi_3 + \left( 6+\frac{\sigma_{1}}{\pi} \right)\phi_{4} - \frac{\sigma_{2}}{4\pi}{\cal T},\label{eq:h00}
\end{eqnarray}
where ${\cal T}$ is a new post-Newtonian potential defined as
\begin{eqnarray}
 {\cal T}(t,\vec{x}) \equiv \int \frac{[\rho(t,\vec{x}')]^{2}}{|x-\vec{x}'|}d^{3}x', \quad       \nabla^{2}{\cal T} = -4\pi\rho^{2}\,.
\end{eqnarray}

With the above, we conclude the expansion of the separable $f(R,T)$ gravity as a function of the PPN potentials. In the next section, we infer the values of the PPN
parameters and compare them with corresponding observational values.

\section{The PPN parameters in $f(R,T)$ gravity}\label{sec:V}

The standard Will-Nordvedt PPN formalism \cite{Will:1993ns, Will:2014kxa} does not include the last term in Eq.~\eqref{eq:h00}. However, the realization $\varphi_2 (T) = \sigma_0 + \sigma_1 T + \sigma_2 T^{2}$ does not keep null the four-divergence of the field equations \eqref{eq:FE2} due to the term $\propto T^{2}$. Then, we assume
\begin{equation}
    \sigma_{2}=0.
\end{equation}

Before we proceed, a cautionary remark seems mandatory. The choice $\sigma_2=0$ is not mandatory. For example, one can carry on the PPN analysis keeping a nonvanishing $\sigma_2$ term. This is a similar to the approach of Ref. \cite{PhysRevD.101.064050} where the PPN parameters in the Palatini $f(R)$ is performed. However, we have checked this would lead to a nontrivial interpretation of the potentials and their connection with the PPN parameters.

With this, and the results obtained in the previous section, the metric up to the first post-Newtonian order can be written as
\begin{eqnarray}
g_{00} = 2U -2U^{2} + \left( 4+\frac{\sigma_{1}}{2\pi} \right)\phi_{1} + \left(4+\frac{2\sigma_{1}}{\pi} \right)\phi_{2} + 2\phi_3 + \left( 6+\frac{\sigma_{1}}{\pi} \right)\phi_{4}\,,\label{eq:g00}
\end{eqnarray}
\begin{eqnarray}
g_{0j} = -\left( \frac{7}{2} +\frac{\sigma_1}{2\pi} \right)V_j - \frac{1}{2}W_j\,,\label{eq:g0j}
\end{eqnarray}
\begin{eqnarray}
g_{ij} = \left[1+ 2\left( 1+ \frac{\sigma_1}{4\pi} \right) U\right]\delta_{ij}\,.\label{eq:gij}
\end{eqnarray}
To extract the PPN parameters from the above metric components we compare it to the Will-Nordvedt general post-Newtonian metric \cite{Will:1993ns}
\begin{align}
g_{00} =& -1 + 2U - 2\beta U^2 + (2 \gamma +2+\alpha_3 +\zeta _1-2 \xi ) \phi_1 
 + 2(3 \gamma -2\beta+1+\zeta _2+ \xi ) \phi_2  \nonumber \\[1ex]
& +2(1+\zeta _3 ) \phi_3 \ + 2(3 \gamma +3\zeta _4-2 \xi ) \phi_4 - (\zeta _1-2 \xi ) {\cal A}  -2\xi \phi_w,  \\[2ex]
g_{0i} =& - \frac{1}{2}(4 \gamma +3+\alpha_1-\alpha_2+ \zeta_1-2\xi) V_i - \frac 1 2(1+\alpha_2- \zeta_1+2\xi) W_i\,, \\[2ex]
g_{ij} =& \ (1+2\gamma \,U)\, \delta_{ij}\,.
\end{align}
From this comparison we obtain the following constraints
\begin{eqnarray}
\beta = 1,\quad \gamma = 1 + \frac{\sigma_1}{4\pi}, \quad \alpha_{1} = \alpha_{2} = \alpha_{3} = 0, \quad \xi = 0, \\
\zeta_{1}=0, \quad \zeta_{2} = \frac{\sigma_{1}}{4\pi}, \quad \zeta_{3} = 0, \quad \zeta_{4} = -\frac{\sigma_{1}}{2\pi}.
\end{eqnarray} 
The observational bounds for all PPN parameters are shown in Table \ref{tab}. The strongest limit for $\sigma_1$ comes from the parameter $\gamma$, which is related to how much curvature is produced per unit rest mass, and states that
\begin{equation}\label{sigma1result}
    |\sigma_1|< 2.9 \times 10^{-4}.
\end{equation}
Since all parameters $\alpha_{i}$ are null, the model does not predict preferred-frame effects. On the other hand, nonvanishing values for the  parameters $\zeta_2$ and $\zeta_4$ indicate a violation of the total momentum conservation. This classifies the model as semiconservative \cite{Will:2014kxa}.

There are well known examples of theories that come from an action and have $\alpha_{1}$ and $\alpha_{2}$ different from zero, but theories with an action are not expected to yield nonzero values for any of the $\zeta$'s and $\alpha_3$ if $\xi=0$ \cite{Lee1685}. A further example of a theory with this characteristic can be seen in \cite{Toniato:2017wmk}.

\begin{table}[ht]
	\centering 
		\begin{tabular}{ c c || c c }
			\toprule
			Parameter & Limit & Parameter & Limit \\
			\hline
			$\gamma-1$& $2.3 \times 10^{-5}$& $\xi$    & $4 \times 10^{-9}$\\
			$\beta-1$ & $8 \times 10^{-5}$ & $\zeta_1$ & $2 \times 10^{-2}$\\
			$\alpha_1$& $4 \times 10^{-5}$ & $\zeta_2$ & $4 \times 10^{-5}$\\ 
			$\alpha_2$& $2 \times 10^{-9}$ & $\zeta_3$ & $1 \times 10^{-8}$\\
			$\alpha_3$& $4 \times 10^{-20}$& $\zeta_4$ & --- \\
			\botrule
		\end{tabular}
    \caption{\label{tab} Limits on the PPN parameters, considering only the strongest limits for each parameter \cite{Will:2014kxa}. The $\zeta_4$ does not have a direct measurement. These limits apply to the absolute value of each parameter.}
\end{table}

\section{Fully conservative model}

The PPN analysis shows that, despite implying $\nabla^{\mu}T_{\mu\nu}=0$ at the field equations level, the model with a linear dependence on $T$ is severely constrained. One can further analyze this by considering the invariance of the theory under diffeomorphisms.

 First, we can consider that the energy-momentum tensor does not depend on derivatives of the matter fields with respect to space-time, as is the case with perfect fluids. Therefore ${\delta f'}/{\delta \Psi} = (\partial f'/\partial\Psi') \delta^{(4)}(x-x')$, and the Eq.~\eqref{eq:matt} becomes
\begin{align}
\frac{\delta S_{\mathrm{m}}}{\delta\Psi}= -\frac{1}{2\kappa^{2}}\frac{\partial T}{\partial \Psi}f_{_T}\sqrt{-g}.\label{eq:matter}
\end{align}
 Imposing that the theory is diffeomorphism-invariant \cite{Wald:1984rg}, one obtains 
\begin{align} 
\int_{\cal M} \left[ (\nabla^{\mu}T_{\mu\nu})\xi^{\nu}\sqrt{-g} + \frac{\delta S_\mathrm{m}}{\delta\Psi}\delta_\xi\Psi \right]d^{4}x=0, \label{eq:diffeo}
\end{align}
where $\delta_\xi \Psi = \xi^{\mu}\partial_{\mu}\Psi$ represents an infinitesimal coordinate change given by the Lie derivative along the vector $\xi^{\mu}$. Plug-in \eqref{eq:matter} and \eqref{eq:diffeo}, we gets
\begin{align}
\nabla^{\nu}T_{\mu\nu} =   \frac{1}{2\kappa^{2}}f_{_T}\nabla_{\mu}T .\label{eq:invdiff}
\end{align}
Therefore, any realization for $f(R,T)$ that leads to $\nabla_{\mu}T^{\mu\nu}=0$ must be such that
\begin{equation}
    f_{_{T}}(R,T)=0.
\end{equation}
Then, $\varphi$ constant is the only solution that simultaneously satisfies \eqref{eq:sol100}, \eqref{eq:sol1ij} and \eqref{eq:invdiff} (with $\nabla_{\mu}T^{\mu\nu}=0$).

\section{Final discussion}

Modified gravity theories can display many distinct new geometrical structures such as, e.g., violation of the usual energy-momentum conservation. This is particularly observed in the cosmological context where the nonconservation leads to a clear signature on the evolution of the pressureless (dark) matter component, the main ingredient for structure formation.
One of the most known example of that is the class of $f(R,T)= f_1(R)+f_2(T)$ theories where $T$ is the trace of the energy momentum tensor. For this theory, assuming a power law dependence $f_2(T) \propto T^n$, the standard cosmological conservation law is recovered when $n=1/2$. We have shown however this is a sub case of a more general notion of conservation in $f(R,T)$ gravity. Only in the expanding cosmological background the solution $f_2(T) \propto T^{1/2}$ can be seen as the full conservative one.

 We then proceed to search for a general form of the $f(R,T)$ function that maintains conservation at the field equations level. Our investigation initially leads us to discover the function $f_2(T)=\sigma_0+\sigma_1 T$. Subsequently, we focus on evaluating this function using solar system data via the Post-Newtonian formalism. As a result, the parameter $\sigma_1$ is found to be considerably limited, as shown in \eqref{sigma1result}. Therefore, we deduce that only the inclusion of a trivial constant term in the gravitational Lagrangian is compatible with both the complete conservation of the energy momentum tensor and solar system tests. This outcome poses a significant challenge to the $T$ dependence of modified gravity theories in practice. Nonetheless, it cannot be contended that this rules out this category of theories since it is possible that the $T$ dependence works on cosmological scales while an auxiliary screening mechanism comes into play to resolve the solar system tests.

It is also important to mention the discussion on possible approaches to $f(R,T)$ gravity with minimal coupling that can be followed in \cite{Fisher:2019ekh,Harko:2020ivb,Fisher:2020zwx}. One can modify the Lagrangian of $f(R,T)$ theories by including Lagrange multipliers that lead to $\nabla_{\mu}T^{\mu\nu}=0$ at the field equations level, as in \cite{Fisher:2019ekh}. On the other hand, one can also investigate which particular realization of $f(R,T)$ agrees with $\nabla_{\mu}T^{\mu\nu}=0$ by keeping the standard gravitational Lagrangian as in this reference and also as argued by Harko and Moraes \cite{Harko:2020ivb} (see also \cite{dosSantos:2018nmu, Pretel:2021kgl}). In any case, the fundamental thermodynamical variables of matter ($\rho$ and $p$) are introduced in $S_m$ and are the same that also appear in the gravitational part of the total Lagrangian, through $T$. Considering the model with minimal coupling, $f(R,T)=f_1(R)+f_2(T)$, including $f_2(T)$ in the matter sector is indeed a technically possible approach, but the new effective thermodynamical variables appearing in the effective matter Lagrangia will have a different meaning comparing to those included in $S_m$, and can be distinguished from the fundamental variables. Imposing $\nabla_{\mu}T^{\mu\nu}=0$ we conclude that the cosmological constant is the favored result, and is part of $f_2(T)$. Including the cosmological constant in the energy density or in the geometric part is a choice, but in any case its value can be unequivocally determined by its physical effects. Therefore, we emphasize that the search for the functional form of $f_2(T)$ and the observational constraints is valid, contrarily to argued in Ref. \cite{Fisher:2019ekh,Fisher:2020zwx} and that the result using the PPN analysis is consistent with the imposition set by Eq. \eqref{eq:invdiff}, for the case of minimal coupling.

\begin{acknowledgments}

The authors thank FAPEMIG/FAPES/CNPq/CAPES for financial support. We thank Junior Toniato for pointing out an important issue on the PPN analysis and Davi Cabral Rodrigues for useful remarks. 
\end{acknowledgments}

\bibliographystyle{apsrev4-1}
\bibliography{MSbib}

\end{document}